\documentclass{article}

\usepackage[nonatbib, final]{neurips_2024_ml4ps}

\usepackage[utf8]{inputenc} 
\DeclareUnicodeCharacter{2264}{$\leq$}
\DeclareUnicodeCharacter{039C}{$\mu$}

\usepackage[T1]{fontenc}    
\usepackage{hyperref}       
\usepackage{url}            
\usepackage{booktabs}       
\usepackage{amsfonts}       
\usepackage{nicefrac}       
\usepackage{microtype}      
\usepackage{xcolor}         
\usepackage{graphicx}

\usepackage{tabularx}

\usepackage[backend=biber]{biblatex}
\title{Using different sources of ground truths and transfer learning to improve the generalization of photometric redshift estimation}

\author{
    Jonathan Soriano$^{1}$ \\
    \texttt{jsoriano@astro.ucla.edu} \\
    \And
    Srinath Saikrishnan$^{1}$ \\
    \texttt{srinathsai22@g.ucla.edu} \\
    \And
    Vikram Seenivasan$^{1}$ \\
    \texttt{vikrams25@g.ucla.edu} \\
    \And
    Bernie Boscoe$^{2}$ \\
    \texttt{boscoeb@sou.edu} \\
    \And
    Jack Singal$^{3}$ \\
    \texttt{jsingal@richmond.edu} \\
    \And
    Tuan Do$^{1}$ \\
    \texttt{tdo@astro.ucla.edu} \\
    \And\\
  $^{1}$ Physics and Astronomy Department, UCLA, Los Angeles, CA 90024 \\
  $^{2}$ Computer Science Department, Southern Oregon University, Ashland, OR 97520 \\
  $^{3}$ Physics Department, University of Richmond, Richmond, VA 23173 \\
}

\begin{document}

\maketitle

\begin{abstract}
In this work, we explore methods to improve galaxy redshift predictions by combining different ground truths. Traditional machine learning models rely on training sets with known spectroscopic redshifts, which are precise but only represent a limited sample of galaxies. To make redshift models more generalizable to the broader galaxy population, we investigate transfer learning and directly combining ground truth redshifts derived from photometry and spectroscopy. We use the COSMOS2020 survey to create a dataset, TransferZ, which includes photometric redshift estimates derived from up to 35 imaging filters using template fitting. This dataset spans a wider range of galaxy types and colors compared to spectroscopic samples, though its redshift estimates are less accurate. We first train a base neural network on TransferZ and then refine it using transfer learning on a dataset of galaxies with more precise spectroscopic redshifts (GalaxiesML). In addition, we train a neural network on a combined dataset of TransferZ and GalaxiesML. Both methods reduce bias by $\sim$ 5x, RMS error by $\sim$ 1.5x, and catastrophic outlier rates by 1.3x on GalaxiesML, compared to a baseline trained only on TransferZ. However, we also find a reduction in performance for RMS and bias when evaluated on TransferZ data. Overall, our results demonstrate these approaches can meet cosmological requirements. 
\end{abstract}

\section{Introduction}

Astronomers are increasingly adopting machine learning methods for redshift estimation, which is crucial for measuring the distances to galaxies in cosmology. Spectroscopic redshifts (spec-z’s) are the most accurate, but they are time-consuming and thus impractical for large-scale surveys involving billions of galaxies \cite[e.g.,][]{ivezic2008, racca2016, breivik2022, euclidcollaboration2024}. Photometric redshifts (photo-z’s) are derived from measurements of the brightness of galaxies (photometry) from images taken at different wavelengths. They are less precise but enable the analysis of much larger datasets \cite{newman2022}. Photometric redshift methods generally fall into two categories: template-fitting and data-driven approaches. In template-fitting \cite[e.g.,][]{ arnouts1999,ilbert2006, brammer2008} a library of broad-band galaxy photometry and redshifts is compared to observed photometry to estimate redshifts. Data-driven methods, often involving machine learning, train models on known redshift samples to predict redshifts for new data \cite[e.g.,][]{bonnett2015a,collister2004, carrasco2015,newman2022,jones2024a}.

A critical factor in the success of machine learning models is the quality and representativeness of the training data. For redshift prediction models, the most accurate training data comes from spectroscopic measurements, which precisely probe emission lines and achieve redshift uncertainties as low as $2\times10^{-4}$ \cite[e.g.,][]{tanaka2018}. However, these measurements are typically limited to bright galaxies with strong emission lines, representing only a small subset of the galaxies in the Universe. The COSMOS2020 survey \cite{weaver2022} offers a broader dataset, covering a wider range of galaxy types. However, the median precision of its redshift measurements is approximately 0.03—about 100 times less precise than spectroscopic redshifts.

The limitation in representativeness highlights the need for methods that can generalize across different types of data. In this paper we explore two approaches of incorporating ground truths from real data for training photometric redshift models: transfer learning and mixing ground truths. Transfer learning \cite{pan2010,weiss2016} offers a promising solution in this regard, allowing models trained on broader, less-precise datasets like COSMOS2020 to be fine tuned on precise but narrower spectroscopic datasets to improve their performance. Mixing ground truths approach is an alternative strategy that combines different sources of redshift measurements at the start of training allowing the model to simultaneously learn from complementary strengths of spectroscopic and photometric redshift datasets. Our approach is novel in its exploration of model generalization by
incorporating different sources of ground truth from real data for training photometric redshift models. Understanding these approaches is particularly important as we look forward to large surveys, where astronomers will need to overcome the gaps in spectroscopic dataset coverage and volume in the initial years. 

\section{Data}

\begin{figure} [htb]
  \centering
  \includegraphics[width=5in]{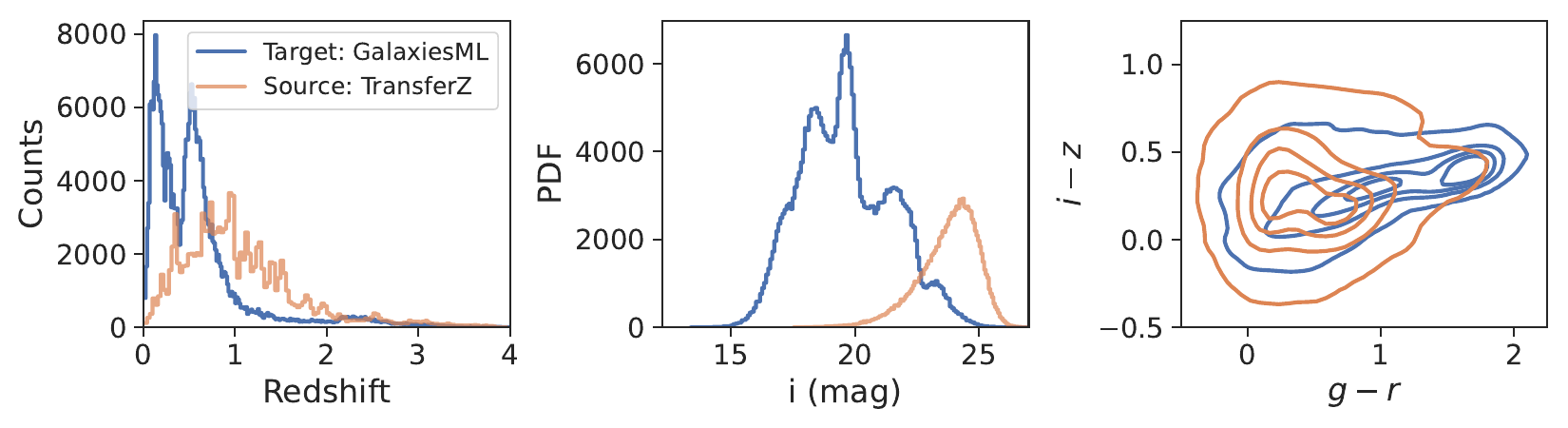}
  \caption{Two datasets: GalaxiesML \cite{do2024} with spectroscopic redshift ground truth and TransferZ with COSMOS2020 survey \cite{weaver2022} multi-band imaging redshift ground truth. The distribution of the dataset in redshift (left), i-band magnitude (center), and color (right) shows how the datasets complement each other to help the model generalize beyond the range of brightness and color sampled by the spectroscopic surveys. }
  \label{fig:data}
\end{figure}

\begin{table}[ht]
  \caption{Data Summary}
  \label{data-table}
  \centering
  \begin{tabular}{l c c c c c}
    \toprule
    Dataset & Number of & Redshift & Median Redshift & i-band mag & No. \\
            & Sources   & 90th percentile    & Uncertainty &  90th percentile  & Filters \\
    \midrule
    TransferZ   & 116,335 & 1.9 & 0.03   & 25 & 5 \\
    GalaxiesML & 286,401 & 1.2 & 0.0002 & 22 & 5 \\
    Combo Data  & 402,408 & 1.5 & 0.0006 & 24 & 5 \\
    \bottomrule
  \end{tabular}
\end{table}

In this work we base our analyses on 5-band photometry to approximate the conditions for the Legacy Survey in Space and Time (LSST), a major upcoming survey \cite{breivik2022, ivezic2008}.  We created the TransferZ dataset by integrating data from two sources: the HSC PDR2 \cite{aihara2019} wide field survey, which provides 5-band \textit{grizy} photometry for a query of 3 million sources, and COSMOS2020 \cite{weaver2022}, which offers up to 35-band photometry for 1.7 million sources along with photometric redshifts. With 35 bands of photometry extending from the ultraviolet to the infrared, the photometric redshifts that can be estimated are much more accurate and precise than those resulting from five-band photometry. This is a reasonable basis for a ground truth redshift value \cite{ilbert2008, singal2022}. From COSMOS2020 we choose redshifts computed using LePhare template fitting \cite{arnouts1999,ilbert2006} with at least 30-band photometry from the CLASSIC subset \cite{weaver2022}. To create TransferZ, we cross-match sources from COSMOS2020 with HSC PDR2 data, filtering for galaxies, and applying quality cuts to ensure reliable ground truth redshifts, resulting in a refined dataset of 116,335 galaxies with 5-band \textit{grizy} photometry (\textit{g}: 4754 \AA, \textit{r}: 6175 \AA, \textit{i}: 7711 \AA, \textit{z}: 8898 \AA, \textit{y}: 9762 \AA) from HSC PDR2 and reliable redshifts from COSMOS2020. For more details, see Appendix \ref{section:dataset}.

We use TransferZ as a broader and more general galaxy sample for redshift estimation to train the baseline model and then transfer learn using GalaxiesML \cite{do2024}, which has ground truth for redshifts from spectroscopy. The two datasets complement each other (Fig. \ref{fig:data}). GalaxiesML has 286,401 galaxies, with 90\% of the galaxies having $\textit{i} < 22$ mag and most galaxies have redshifts $<1.2$ (note that larger magnitude values mean the galaxies are fainter). This dataset is built on HSC PDR2 \cite{aihara2019} and its associated spectroscopic database \cite{lilly2009,bradshaw2013,mclure2012,skelton2014,momcheva2016,lefevre2013,garilli2014,liske2015,davis2003,newman2013,coil2011,cool2013}. TransferZ has 90\% of its galaxies with $i < 25$ mag and redshifts $<1.9$. TransferZ contains a higher number of galaxies in the cosmologically relevant range of $0.3 < z < 1.5$, potentially enabling representational analysis at higher redshifts than GalaxiesML alone. While TransferZ probes much fainter galaxies and more galaxy types, the redshift uncertainties from the 35-band photometry are typically 100 times larger than GalaxiesML with spectroscopy (Table \ref{data-table}). We note that there are 500 galaxies (about $0.1\%$ of the total) in common between both TransferZ and GalaxiesML (Fig. \ref{fig:data}). We assume the impact of this overlap is negligible for this experiment, but this can be verified in the future. 

We also created a combination dataset called Combo that combines both TransferZ and GalaxiesML to test whether combining two types of ground truth is equivalent to transfer learning from one dataset to another. When there are both spectroscopic and COSMOS2020 photometric redshifts for the same galaxy, we choose to include only the spectroscopic redshift, because it is more accurate (about 500 galaxies are affected). The combo dataset consists of 402,408 galaxies. The datasets are split into 80\% training, 10\% validation, and 10\% testing sets. In the following sections, we refer to TransferZ as the source data, GalaxiesML as the target data, and Combo as combo data. TransferZ is made available on Zenodo with a DOI:  10.5281/zenodo.14218996.

\section{Methodology \& metrics}

We employed a neural network (NN) architecture based on \cite{jones2022, jones2024a} for photometric redshift estimation, consisting of four fully connected layers with ReLU activation and a skip connection. Hyperparameter tuning was performed with the source training and validation data using HyperBand \cite{li2018}. The Hyperband search space for training the NN includes 1 to 10 hidden layers with 32 to 2048 neurons per layer, whether to include a skip connection, and whether to add additional dense layers. If additional dense layers are included, they range from 1 to 10 hidden layers with 32 to 4096 neurons per layer. The final model has the four initial hidden layers with 200 neurons each followed by a skip connection and two additional hidden layers with 2000 neurons each. All hidden layers use the rectified linear unit (ReLU) activation function. 

The base model (NN-Base) was trained on the TransferZ dataset using the Adam optimizer with a learning rate of \(5 \times 10^{-4}\), a batch size of 512, and for 500 epochs. Transfer learning (NN-TL) was then applied by fine-tuning the NN-Base model on the GalaxiesML dataset, freezing all layers except the input and the first and fifth dense layers, with a reduced learning rate of \(5 \times 10^{-10}\) and trained for 1000 epochs.

The model trained on the combined data (NN-Combo) is hyperparameter optimized similarly to the NN-Base model. The final model has 6 hidden layers and a skip connection between inputs and the hidden layers. NN-Combo was trained on the Combo dataset using the Adam optimizer with a learning rate of \(5 \times 10^{-4}\), a batch size of 512, and for 2000 epochs. The three model trainings achieve optimal performance before the reported epochs since learning curves plateau earlier. 

A custom loss function, $L(\Delta z) = 1 - \frac{1}{1 + (\Delta z/0.15)^2}$, is used \cite{tanaka2018} for training. The photometry data is normalized separately for each training stage. Performance was evaluated using bias, root mean square error (RMS), and the catastrophic outlier rate on the test sets within the redshift range of $0.3 < z < 1.5$. The bias metric is the median of the bias distribution defined as $b = (z_{photo} - z_{truth})/(1 + z_{truth})$ where $z_{photo}$ and $z_{truth}$ is the estimated photometric redshift and the ground truth redshift, respectively. The RMS is defined as the interquartile range of the bias distribution divided by 1.349 weighted by the median redshift in the bin ($1+\overline{z_{truth}}$) \cite{graham2020}. This definition of RMS is less sensitive to outlier rates than standard definitions. The catastrophic outlier rate is defined as the fraction of objects where the absolute difference between photometric and true redshifts exceeds 1.0, expressed as $|z_{photo} - z_{truth}| > 1.0$. These are among the most important metrics for cosmology \cite{blake2005, ivezic2008}. The metrics are evaluated for NN-Base, NN-TL, and NN-Combo test datasets. We compare our metrics to those in \cite{jones2024}, where they use a NN trained on GalaxiesML. The comparison highlights the benefits of mixing ground truths against a spectroscopic ground truth.  

\section{Results \& Discussion}

In this work, we test different ways of combining different sources of ground truth for photometric redshifts - either through transfer learning or by combining the training datasets. We find that both methods are better than the base model, which is only trained on the TransferZ dataset. This suggests that it is possible to improve photometric redshift estimates by combing multiple sources of ground truth. The choice between the two methods should be based on which metric best serves the scientific objectives.  Below we summarize the benefits and limitations of our approaches: transfer learning of the NN model on the source (TransferZ) and target (GalaxiesML) datasets, and a model trained on the combination of GalaxiesML and TransferZ (Combo) dataset. 

\begin{figure}[htb]
    \centering
    \includegraphics[width=5in]{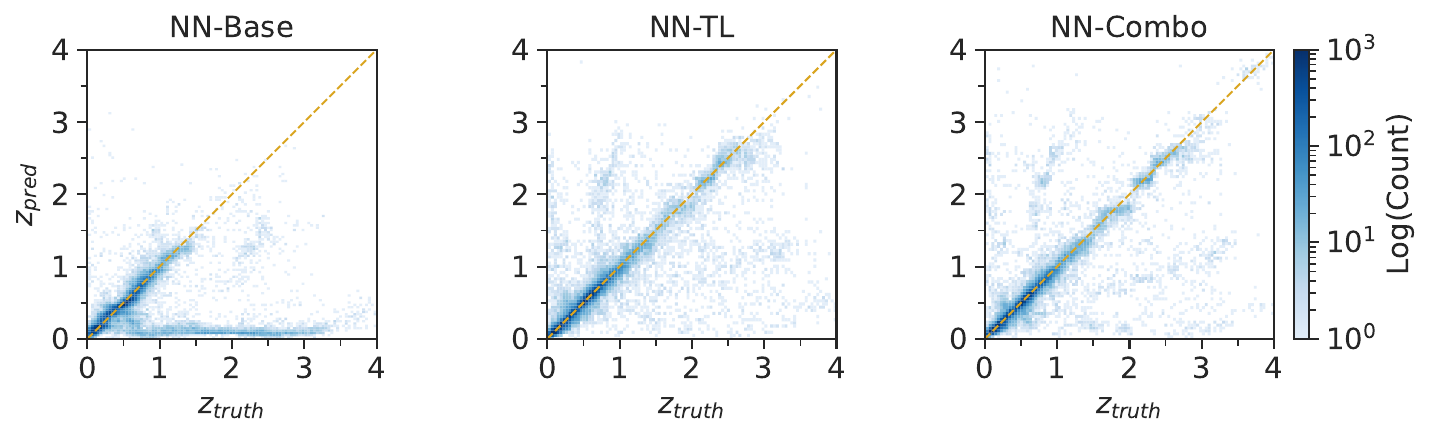}
    \caption{Comparison of redshift predictions from the three neural network models in this work (NN-Base, NN-TL, and NN-Combo) against true redshift values. We show results for the GalaxiesML test dataset using spectroscopic redshift as ground truth. Results shown are from one randomly selected run out of 100 total iterations.}
    \label{fig:prediction_plot}
\end{figure}

\begin{table}[htb]
 \caption{Model Metrics Summary}
 \label{model-performance-table}
 \centering
 \begin{tabular}{llrrr}
   \toprule
   \multicolumn{2}{c}{} & \multicolumn{1}{c}{Bias ($\times 10^{-3})$} & \multicolumn{1}{c}{RMS ($\times 10^{-3}$)} & \multicolumn{1}{c}{Cat. Outlier} \\
   \multicolumn{4}{c}{} & \multicolumn{1}{c}{Rate ($\times 10^{-2}$)}  \\
   \midrule
   NN-Base & TransferZ  & -0.69 $\pm$ 2.37 & 22.6 $\pm$ 0.35 & 1.29 $\pm$ 0.25 \\
           & GalaxiesML & -10.9 $\pm$ 5.71 & 22.4 $\pm$ 1.23 & 2.49 $\pm$ 0.15 \\
   NN-TL  & TransferZ   &  7.45 $\pm$ 3.12 & 28.5 $\pm$ 0.86 & 0.40 $\pm$ 0.13 \\
           & GalaxiesML & -1.51 $\pm$ 1.66 & 15.4 $\pm$ 0.27 & 1.68 $\pm$ 0.14 \\
   NN-Combo & TransferZ &  1.14 $\pm$ 1.74 & 23.0 $\pm$ 0.33 & 1.33 $\pm$ 0.29 \\
           & GalaxiesML & -1.92 $\pm$ 1.68 & 15.0 $\pm$ 0.23 & 1.89 $\pm$ 0.17 \\
   \bottomrule
 \end{tabular}
\end{table}

Model predictions are evaluated against test datasets comprising of 40,914 galaxies from GalaxiesML and 11,633 from TransferZ. Both NN-TL and NN-Combo perform comparably within the redshift range of $0.3\leq z \leq 1.5$, exhibiting similar prediction patterns and improving upon the base model (Fig. \ref{fig:prediction_plot}). Both NN-TL and NN-Combo have small scatter in their predictions at these redshift ranges compared to the base model. Quantitatively, we evaluate the model's predictions using three metrics -- bias, RMS, and catastrophic outlier rate, these are reported in Table \ref{model-performance-table}. We compare the fractional change of the metrics on both target and source data against NN-Base metrics. Evaluating NN-TL (NN-Combo) on the target data shows $7.19$ ($5.65$) times lower bias, $1.45$ ($1.49$) times lower RMS, and $1.48$ ($1.32$) times lower catastrophic outlier rate. The largest magnitude change can be seen for bias, which is important as confidence in a redshift bin assignment is crucial for cosmological measurements requiring precise redshift distributions. The catastrophic outlier rate remains below 10\% across all models, showing that neural network can effectively control a major systematic contaminant in photometric redshift measurements. 

For the source data metrics, we find NN-TL reduces the catastrophic outlier rate compared to NN-Base, but performs worse on bias and RMS. The metrics of NN-Combo are comparable to those of NN-Base on the source data. Evaluating NN-TL (NN-Combo) on the source data shows 3.24 times lower (1.23 times higher) catastrophic outlier rate. However, it shows a bias $10.7$ ($1.64$) times higher and RMS $1.26$ ($1.02$) times higher. The increase in bias on the source data after transfer learning on a new ground truth suggests some loss of the originally learned features. The model trained on combined ground truths is similarly affected but to a lesser extent, meaning it better leverages both sources of truth.

\begin{figure}[htb]
  \centering
  \includegraphics[width=5in]{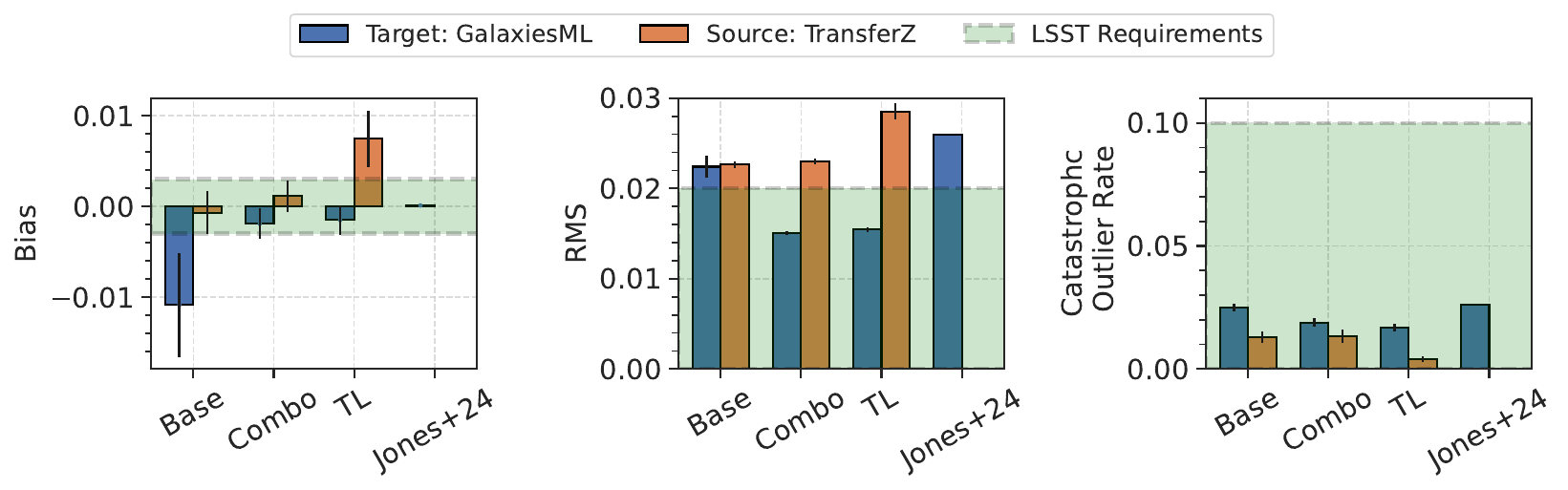}
  \caption{From left to right, comparison of the bias, outlier and RMS metrics between the baseline NN, transfer-learnt NN, combo NN, and \cite{jones2024}. The metrics are evaluated on the target (blue) and source (orange) data within the range of $0.3\leq z \leq 1.5$. The error bars are generated from 100 random initializations of the model training. We report \cite{jones2024} scatter value for RMS. While \cite{jones2024} use a different RMS definition, our RMS calculation is equivalent to their reported scatter value.}
  \label{fig:merticsbarplot}
\end{figure}

The results also show both transfer learning approaches and models trained on combined ground truth data improve photometric redshift predictions compared to single-dataset training. Our transfer learning model performs better than the one from \cite{jones2024} (henceforth J24), which uses only the target dataset (Fig. \ref{fig:merticsbarplot}). In comparing our two approaches for handling mixed ground truth data, the advantage of the models depends on the metric. The Combo model performs better than the transfer learn model on bias (6 times lower) and RMS (1.2 times lower) on target data, and marginally on RMS (1.03 times lower) for source. However, it performs worse on bias (1.27 times higher) on the target dataset. Additionally, the catastrophic outlier rate on target data is  $1.12$ times higher and on source data $3.24$ times higher. Depending on the science needs, the best approach needs to align with the science goal.

There are several limitations to both NN-TL and NN-Combo. The model from J24 achieved a bias of one order magnitude lower than our approach in NN-TL and NN-Combo evaluated on the GalaxiesML.  In addition, there are some notable features in the figure showing the prediction versus true redshift, such as the clustered predictions at $0.5 \leq z_{truth} \leq 1$ and $1.5 \leq z_{pred} \leq 3$ in the target dataset (Fig. \ref{fig:prediction_plot}). This suggests there is a source of systematic error (to be analyzed in a forthcoming paper). NN-TL also shows limited accuracy at $z_{truth}>3$ from sparse training data in these redshift ranges at each training step, while NN-Combo achieves accurate predictions through its training approach of combining both datasets in one training step. 

Previous work explored the use of transfer learning in the context of improving redshift estimates using a combination of simulated and real data \cite{eriksen2020a,moskowitz2024}, but to our knowledge, this is the first work to apply transfer learning from only real data to generalize photometric redshift predictions. Both of our models (NN-TL and NN-Combo) are capable of incorporating two sources of ground truth, but there are a number of ways that the models can be improved upon. For example, our models do not produce uncertainties. Extending transfer learning to probabilistic ML models will be important for more scientific applications \cite[e.g,][]{benitez2000,pasquet2019,treyer2023,jones2024a}. Additional tests of how well this model generalizes will be important to validate the improvements. One approach could be to use this model to detect galaxy clusters in existing data. Galaxies within the same cluster will be at the same redshift, but will consist of many galaxy types, which allows us to independently quantify the generalizability of the models. Ultimately, this could help solve one of the most challenging problems for surveys like LSST: providing reliable photometric redshift measurements for testing models of cosmology.

\clearpage

\printbibliography

@article{tanaka2018,
  title = {Photometric Redshifts for {{Hyper Suprime-Cam Subaru Strategic Program Data Release}} 1},
  author = {Tanaka, Masayuki and Coupon, Jean and Hsieh, Bau-Ching and Mineo, Sogo and Nishizawa, Atsushi J and Speagle, Joshua and Furusawa, Hisanori and Miyazaki, Satoshi and Murayama, Hitoshi},
  year = {2018},
  month = jan,
  journal = {Publications of the Astronomical Society of Japan},
  volume = {70},
  number = {SP1},
  pages = {S9},
  issn = {0004-6264},
  doi = {10.1093/pasj/psx077}
}

@article{scoville2007,
  title = {The {{Cosmic Evolution Survey}} ({{COSMOS}}): {{Overview}}},
  shorttitle = {The {{Cosmic Evolution Survey}} ({{COSMOS}})},
  author = {Scoville, N. and Aussel, H. and Brusa, M. and Capak, P. and Carollo, C. M. and Elvis, M. and Giavalisco, M. and Guzzo, L. and Hasinger, G. and Impey, C. and Kneib, J. -P. and LeFevre, O. and Lilly, S. J. and Mobasher, B. and Renzini, A. and Rich, R. M. and Sanders, D. B. and Schinnerer, E. and Schminovich, D. and Shopbell, P. and Taniguchi, Y. and Tyson, N. D.},
  year = {2007},
  month = sep,
  journal = {The Astrophysical Journal Supplement Series},
  volume = {172},
  pages = {1--8},
  issn = {0067-0049},
  doi = {10.1086/516585}
}

@article{pasquet2019,
  title = {Photometric Redshifts from {{SDSS}} Images Using a Convolutional Neural Network},
  author = {Pasquet, Johanna and Bertin, E. and Treyer, M. and Arnouts, S. and Fouchez, D.},
  year = {2019},
  month = jan,
  journal = {Astronomy \& Astrophysics},
  volume = {621},
  pages = {A26},
  issn = {0004-6361, 1432-0746},
  doi = {10.1051/0004-6361/201833617}
}

@article{newman2022,
  title = {Photometric {{Redshifts}} for {{Next-Generation Surveys}}},
  author = {Newman, Jeffrey A. and Gruen, Daniel},
  year = {2022},
  journal = {Annual Review of Astronomy and Astrophysics},
  volume = {60},
  number = {1},
  pages = {363--414},
  doi = {10.1146/annurev-astro-032122-014611}
}

@article{lefevre2013,
  title = {The {{VIMOS VLT Deep Survey}} Final Data Release: A Spectroscopic Sample of 35 016 Galaxies and {{AGN}} out to z \textasciitilde{} 6.7 Selected with 17.5 ≤ i{{{\textsubscript{AB}}}} ≤ 24.75},
  shorttitle = {The {{VIMOS VLT Deep Survey}} Final Data Release},
  author = {Le Fèvre, O. and Cassata, P. and Cucciati, O. and Garilli, B. and Ilbert, O. and Le Brun, V. and Maccagni, D. and Moreau, C. and Scodeggio, M. and Tresse, L. and Zamorani, G. and Adami, C. and Arnouts, S. and Bardelli, S. and Bolzonella, M. and Bondi, M. and Bongiorno, A. and Bottini, D. and Cappi, A. and Charlot, S. and Ciliegi, P. and Contini, T. and {de la Torre}, S. and Foucaud, S. and Franzetti, P. and Gavignaud, I. and Guzzo, L. and Iovino, A. and Lemaux, B. and {López-Sanjuan}, C. and McCracken, H. J. and Marano, B. and Marinoni, C. and Mazure, A. and Mellier, Y. and Merighi, R. and Merluzzi, P. and Paltani, S. and Pellò, R. and Pollo, A. and Pozzetti, L. and Scaramella, R. and Tasca, L. and Vergani, D. and Vettolani, G. and Zanichelli, A. and Zucca, E.},
  year = {2013},
  month = nov,
  journal = {Astronomy and Astrophysics},
  volume = {559},
  pages = {A14},
  issn = {0004-6361},
  doi = {10.1051/0004-6361/201322179},
  language = {en}
}

@misc{euclidcollaboration2024,
  title = {Euclid. {{I}}. {{Overview}} of the {{Euclid}} Mission},
  author = {Euclid Collaboration and Mellier, Y. and Abdurro'uf and Barroso, J. A. Acevedo and Achúcarro, A. and Adamek, J. and Adam, R. and Addison, G. E. and Aghanim, N. and Aguena, M. and Ajani, V. and Akrami, Y. and {Al-Bahlawan}, A. and Alavi, A. and Albuquerque, I. S. and Alestas, G. and Alguero, G. and Allaoui, A. and Allen, S. W. and Allevato, V. and {Alonso-Tetilla}, A. V. and Altieri, B. and {Alvarez-Candal}, A. and Amara, A. and Amendola, L. and Amiaux, J. and Andika, I. T. and Andreon, S. and Andrews, A. and Angora, G. and Angulo, R. E. and Annibali, F. and Anselmi, A. and Anselmi, S. and Arcari, S. and Archidiacono, M. and Aricò, G. and Arnaud, M. and Arnouts, S. and Asgari, M. and Asorey, J. and Atayde, L. and Atek, H. and {Atrio-Barandela}, F. and Aubert, M. and Aubourg, E. and Auphan, T. and Auricchio, N. and Aussel, B. and Aussel, H. and Avelino, P. P. and Avgoustidis, A. and Avila, S. and Awan, S. and Azzollini, R. and Baccigalupi, C. and Bachelet, E. and Bacon, D. and Baes, M. and Bagley, M. B. and {Bahr-Kalus}, B. and {Balaguera-Antolinez}, A. and Balbinot, E. and Balcells, M. and Baldi, M. and Baldry, I. and Balestra, A. and Ballardini, M. and Ballester, O. and Balogh, M. and Bañados, E. and Barbier, R. and Bardelli, S. and Barreiro, T. and Barriere, J.-C. and Barros, B. J. and Barthelemy, A. and Bartolo, N. and Basset, A. and Battaglia, P. and Battisti, A. J. and Baugh, C. M. and Baumont, L. and Bazzanini, L. and Beaulieu, J.-P. and Beckmann, V. and Belikov, A. N. and Bel, J. and Bellagamba, F. and Bella, M. and Bellini, E. and Benabed, K. and Bender, R. and Benevento, G. and Bennett, C. L. and Benson, K. and Bergamini, P. and {Bermejo-Climent}, J. R. and Bernardeau, F. and Bertacca, D. and Berthe, M. and Berthier, J. and Bethermin, M. and Beutler, F. and Bevillon, C. and Bhargava, S. and Bhatawdekar, R. and Bisigello, L. and Biviano, A. and Blake, R. P. and Blanchard, A. and Blazek, J. and Blot, L. and Bosco, A. and Bodendorf, C. and Boenke, T. and Böhringer, H. and Bolzonella, M. and Bonchi, A. and Bonici, M. and Bonino, D. and Bonino, L. and Bonvin, C. and Bon, W. and Booth, J. T. and Borgani, S. and Borlaff, A. S. and Borsato, E. and Bosco, A. and Bose, B. and Botticella, M. T. and Boucaud, A. and Bouche, F. and Boucher, J. S. and Boutigny, D. and Bouvard, T. and Bouy, H. and Bowler, R. A. A. and Bozza, V. and Bozzo, E. and Branchini, E. and {Brau-Nogue}, S. and Brekke, P. and Bremer, M. N. and Brescia, M. and Breton, M.-A. and Brinchmann, J. and Brinckmann, T. and {Brockley-Blatt}, C. and Brodwin, M. and Brouard, L. and Brown, M. L. and Bruton, S. and Bucko, J. and Buddelmeijer, H. and Buenadicha, G. and Buitrago, F. and Burger, P. and Burigana, C. and Busillo, V. and Busonero, D. and Cabanac, R. and {Cabayol-Garcia}, L. and Cagliari, M. S. and Caillat, A. and Caillat, L. and Calabrese, M. and Calabro, A. and Calderone, G. and Calura, F. and Quevedo, B. Camacho and Camera, S. and Campos, L. and {Canas-Herrera}, G. and Candini, G. P. and Cantiello, M. and Capobianco, V. and Cappellaro, E. and Cappelluti, N. and Cappi, A. and Caputi, K. I. and Cara, C. and Carbone, C. and Cardone, V. F. and Carella, E. and Carlberg, R. G. and Carle, M. and Carminati, L. and Caro, F. and Carrasco, J. M. and Carretero, J. and Carrilho, P. and Duque, J. Carron and Carry, B. and Carvalho, A. and Carvalho, C. S. and Casas, R. and Casas, S. and Casenove, P. and Casey, C. M. and Cassata, P. and Castander, F. J. and Castelao, D. and Castellano, M. and Castiblanco, L. and Castignani, G. and Castro, T. and Cavet, C. and Cavuoti, S. and Chabaud, P.-Y. and Chambers, K. C. and Charles, Y. and Charlot, S. and Chartab, N. and Chary, R. and Chaumeil, F. and Cho, H. and Chon, G. and Ciancetta, E. and Ciliegi, P. and Cimatti, A. and Cimino, M. and Cioni, M.-R. L. and Claydon, R. and Cleland, C. and Clément, B. and Clements, D. L. and Clerc, N. and Clesse, S. and Codis, S. and Cogato, F. and Colbert, J. and Cole, R. E. and Coles, P. and Collett, T. E. and Collins, R. S. and {Colodro-Conde}, C. and Colombo, C. and Combes, F. and Conforti, V. and Congedo, G. and Conseil, S. and Conselice, C. J. and Contarini, S. and Contini, T. and Conversi, L. and Cooray, A. R. and Copin, Y. and Corasaniti, P.-S. and {Corcho-Caballero}, P. and Corcione, L. and Cordes, O. and Corpace, O. and Correnti, M. and Costanzi, M. and Costille, A. and Courbin, F. and Mifsud, L. Courcoult and Courtois, H. M. and Cousinou, M.-C. and Covone, G. and Cowell, T. and Cragg, C. and Cresci, G. and Cristiani, S. and Crocce, M. and Cropper, M. and Crouzet, P. E. and Csizi, B. and Cuby, J.-G. and Cucchetti, E. and Cucciati, O. and Cuillandre, J.-C. and Cunha, P. A. C. and Cuozzo, V. and Daddi, E. and D'Addona, M. and Dafonte, C. and Dagoneau, N. and Dalessandro, E. and Dalton, G. B. and D'Amico, G. and Dannerbauer, H. and Danto, P. and Das, I. and Da Silva, A. and {da Silva}, R. and Daste, G. and Davies, J. E. and Davini, S. and {de Boer}, T. and Decarli, R. and De Caro, B. and Degaudenzi, H. and Degni, G. and {de Jong}, J. T. A. and {de la Bella}, L. F. and {de la Torre}, S. and Delhaise, F. and Delley, D. and Delucchi, G. and De Lucia, G. and Denniston, J. and De Paolis, F. and De Petris, M. and Derosa, A. and Desai, S. and Desjacques, V. and Despali, G. and Desprez, G. and {De Vicente-Albendea}, J. and Deville, Y. and Dias, J. D. F. and {Díaz-Sánchez}, A. and Diaz, J. J. and Di Domizio, S. and Diego, J. M. and Di Ferdinando, D. and Di Giorgio, A. M. and Dimauro, P. and Dinis, J. and Dolag, K. and Dolding, C. and Dole, H. and Sánchez, H. Domínguez and Doré, O. and Dournac, F. and Douspis, M. and Dreihahn, H. and Droge, B. and Dryer, B. and Dubath, F. and Duc, P.-A. and Ducret, F. and Duffy, C. and Dufresne, F. and Duncan, C. A. J. and Dupac, X. and Duret, V. and Durrer, R. and Durret, F. and Dusini, S. and Ealet, A. and Eggemeier, A. and Eisenhardt, P. R. M. and Elbaz, D. and Elkhashab, M. Y. and Ellien, A. and Endicott, J. and Enia, A. and Erben, T. and Vigo, J. A. Escartin and Escoffier, S. and Sanz, I. Escudero and Essert, J. and Ettori, S. and Ezziati, M. and Fabbian, G. and Fabricius, M. and Fang, Y. and Farina, A. and Farina, M. and Farinelli, R. and Farrens, S. and Faustini, F. and Feltre, A. and Ferguson, A. M. N. and Ferrando, P. and Ferrari, A. G. and {Ferré-Mateu}, A. and Ferreira, P. G. and Ferreras, I. and Ferrero, I. and Ferriol, S. and Ferruit, P. and Filleul, D. and Finelli, F. and Finkelstein, S. L. and Finoguenov, A. and Fiorini, B. and Flentge, F. and Focardi, P. and Fonseca, J. and Fontana, A. and Fontanot, F. and Fornari, F. and Fosalba, P. and Fossati, M. and Fotopoulou, S. and Fouchez, D. and Fourmanoit, N. and Frailis, M. and {Fraix-Burnet}, D. and Franceschi, E. and Franco, A. and Franzetti, P. and Freihoefer, J. and Frittoli, G. and Frugier, P.-A. and Frusciante, N. and Fumagalli, A. and Fumagalli, M. and Fumana, M. and Fu, Y. and Gabarra, L. and Galeotta, S. and Galluccio, L. and Ganga, K. and Gao, H. and {García-Bellido}, J. and Garcia, K. and Gardner, J. P. and Garilli, B. and {Gaspar-Venancio}, L.-M. and Gasparetto, T. and Gautard, V. and Gavazzi, R. and Gaztanaga, E. and Genolet, L. and Santos, R. Genova and Gentile, F. and George, K. and Ghaffari, Z. and Giacomini, F. and Gianotti, F. and Gibb, G. P. S. and Gillard, W. and Gillis, B. and Ginolfi, M. and Giocoli, C. and Girardi, M. and Giri, S. K. and Goh, L. W. K. and {Gómez-Alvarez}, P. and Gonzalez, A. H. and Gonzalez, E. J. and Gonzalez, J. C. and Beauchamps, S. Gouyou and Gozaliasl, G. and {Gracia-Carpio}, J. and Grandis, S. and Granett, B. R. and Granvik, M. and Grazian, A. and Gregorio, A. and Grenet, C. and Grillo, C. and Grupp, F. and Gruppioni, C. and Gruppuso, A. and Guerbuez, C. and Guerrini, S. and Guidi, M. and Guillard, P. and Gutierrez, C. M. and Guttridge, P. and Guzzo, L. and Gwyn, S. and Haapala, J. and Haase, J. and Haddow, C. R. and Hailey, M. and Hall, A. and Hall, D. and Hamaus, N. and Haridasu, B. S. and {Harnois-Déraps}, J. and Harper, C. and Hartley, W. G. and Hasinger, G. and Hassani, F. and Hatch, N. A. and Haugan, S. V. H. and Häußler, B. and Heavens, A. and Heisenberg, L. and Helmi, A. and Helou, G. and Hemmati, S. and Henares, K. and Herent, O. and {Hernández-Monteagudo}, C. and Heuberger, T. and Hewett, P. C. and Heydenreich, S. and Hildebrandt, H. and Hirschmann, M. and Hjorth, J. and Hoar, J. and Hoekstra, H. and Holland, A. D. and Holliman, M. S. and Holmes, W. and Hook, I. and Horeau, B. and Hormuth, F. and Hornstrup, A. and Hosseini, S. and Hu, D. and Hudelot, P. and Hudson, M. J. and {Huertas-Company}, M. and Huff, E. M. and Hughes, A. C. N. and Humphrey, A. and Hunt, L. K. and Huynh, D. D. and Ibata, R. and Ichikawa, K. and {Iglesias-Groth}, S. and Ilbert, O. and Ilić, S. and Ingoglia, L. and Iodice, E. and Israel, H. and Israelsson, U. E. and Izzo, L. and Jablonka, P. and Jackson, N. and Jacobson, J. and Jafariyazani, M. and Jahnke, K. and Jansen, H. and Jarvis, M. J. and Jasche, J. and Jauzac, M. and Jeffrey, N. and Jhabvala, M. and {Jimenez-Teja}, Y. and Muñoz, A. Jimenez and Joachimi, B. and Johansson, P. H. and Joudaki, S. and Jullo, E. and Kajava, J. J. E. and Kang, Y. and Kannawadi, A. and Kansal, V. and Karagiannis, D. and Kärcher, M. and Kashlinsky, A. and Kazandjian, M. V. and Keck, F. and Keihänen, E. and Kerins, E. and Kermiche, S. and Khalil, A. and Kiessling, A. and Kiiveri, K. and Kilbinger, M. and Kim, J. and King, R. and Kirkpatrick, C. C. and Kitching, T. and Kluge, M. and Knabenhans, M. and Knapen, J. H. and Knebe, A. and Kneib, J.-P. and Kohley, R. and Koopmans, L. V. E. and Koskinen, H. and Koulouridis, E. and Kou, R. and Kovács, A. and Kova\{č\}ić, I. and Kowalczyk, A. and Koyama, K. and Kraljic, K. and Krause, O. and Kruk, S. and Kubik, B. and Kuchner, U. and Kuijken, K. and Kümmel, M. and Kunz, M. and {Kurki-Suonio}, H. and Lacasa, F. and Lacey, C. G. and La Franca, F. and Lagarde, N. and Lahav, O. and Laigle, C. and La Marca, A. and La Marle, O. and Lamine, B. and Lam, M. C. and Lançon, A. and Landt, H. and Langer, M. and Lapi, A. and Larcheveque, C. and Larsen, S. S. and Lattanzi, M. and Laudisio, F. and Laugier, D. and Laureijs, R. and Lavaux, G. and Lawrenson, A. and Lazanu, A. and Lazeyras, T. and Boulc'h, Q. Le and Brun, A. M. C. Le and Brun, V. Le and Leclercq, F. and Lee, S. and Graet, J. Le and Legrand, L. and Leirvik, K. N. and Jeune, M. Le and Lembo, M. and Mignant, D. Le and Lepinzan, M. D. and Lepori, F. and Lesci, G. F. and Lesgourgues, J. and Leuzzi, L. and Levi, M. E. and Liaudat, T. I. and Libet, G. and Liebing, P. and Ligori, S. and Lilje, P. B. and Lin, C.-C. and Linde, D. and Linder, E. and Lindholm, V. and Linke, L. and Li, S.-S. and Liu, S. J. and Lloro, I. and Lobo, F. S. N. and Lodieu, N. and Lombardi, M. and Lombriser, L. and Lonare, P. and Longo, G. and {López-Caniego}, M. and Lopez, X. Lopez and Alvarez, J. Lorenzo and Loureiro, A. and Loveday, J. and Lusso, E. and {Macias-Perez}, J. and Maciaszek, T. and Magliocchetti, M. and Magnard, F. and Magnier, E. A. and Magro, A. and Mahler, G. and Mainetti, G. and Maino, D. and Maiorano, E. and Maiorano, E. and Malavasi, N. and Mamon, G. A. and Mancini, C. and Mandelbaum, R. and Manera, M. and {Manjón-García}, A. and Mannucci, F. and Mansutti, O. and Outeiro, M. Manteiga and Maoli, R. and Maraston, C. and Marcin, S. and {Marcos-Arenal}, P. and {Margalef-Bentabol}, B. and Marggraf, O. and Marinucci, D. and Marinucci, M. and Markovic, K. and Marleau, F. R. and Marpaud, J. and Martignac, J. and {Martín-Fleitas}, J. and {Martin-Moruno}, P. and Martin, E. L. and Martinelli, M. and Martinet, N. and Martin, H. and Martins, C. J. A. P. and Marulli, F. and Massari, D. and Massey, R. and Masters, D. C. and Matarrese, S. and Matsuoka, Y. and Matthew, S. and Maughan, B. J. and Mauri, N. and Maurin, L. and Maurogordato, S. and McCarthy, K. and McConnachie, A. W. and McCracken, H. J. and McDonald, I. and McEwen, J. D. and McPartland, C. J. R. and Medinaceli, E. and Mehta, V. and Mei, S. and Melchior, M. and Melin, J.-B. and Ménard, B. and Mendes, J. and {Mendez-Abreu}, J. and Meneghetti, M. and Mercurio, A. and Merlin, E. and Metcalf, R. B. and Meylan, G. and Migliaccio, M. and Mignoli, M. and Miller, L. and Miluzio, M. and {Milvang-Jensen}, B. and Mimoso, J. P. and Miquel, R. and Miyatake, H. and Mobasher, B. and Mohr, J. J. and Monaco, P. and Monguió, M. and Montoro, A. and Mora, A. and Dizgah, A. Moradinezhad and Moresco, M. and Moretti, C. and Morgante, G. and Morisset, N. and Moriya, T. J. and Morris, P. W. and Mortlock, D. J. and Moscardini, L. and Mota, D. F. and Moustakas, L. A. and Moutard, T. and Müller, T. and Munari, E. and Murphree, G. and Murray, C. and Murray, N. and Musi, P. and Nadathur, S. and Nagam, B. C. and Nagao, T. and Naidoo, K. and Nakajima, R. and Nally, C. and Natoli, P. and {Navarro-Alsina}, A. and Girones, D. Navarro and Neissner, C. and Nersesian, A. and Nesseris, S. and {Nguyen-Kim}, H. N. and Nicastro, L. and Nichol, R. C. and Nielbock, M. and Niemi, S.-M. and Nieto, S. and Nilsson, K. and Noller, J. and Norberg, P. and Nourizonoz, A. and Ntelis, P. and Nucita, A. A. and Nugent, P. and Nunes, N. J. and Nutma, T. and Ocampo, I. and Odier, J. and Oesch, P. A. and Oguri, M. and Oliveira, D. Magalhaes and Onoue, M. and Oosterbroek, T. and Oppizzi, F. and Ordenovic, C. and Osato, K. and Pacaud, F. and Pace, F. and Padilla, C. and Paech, K. and Pagano, L. and Page, M. J. and Palazzi, E. and Paltani, S. and Pamuk, S. and Pandolfi, S. and Paoletti, D. and Paolillo, M. and Papaderos, P. and Pardede, K. and Parimbelli, G. and Parmar, A. and Partmann, C. and Pasian, F. and Passalacqua, F. and Paterson, K. and Patrizii, L. and Pattison, C. and {Paulino-Afonso}, A. and Paviot, R. and Peacock, J. A. and Pearce, F. R. and Pedersen, K. and Peel, A. and Peletier, R. F. and Ibanez, M. Pellejero and Pello, R. and Penny, M. T. and Percival, W. J. and {Perez-Garrido}, A. and Perotto, L. and Pettorino, V. and Pezzotta, A. and Pezzuto, S. and Philippon, A. and Piersanti, O. and Pietroni, M. and Piga, L. and Pilo, L. and Pires, S. and Pisani, A. and Pizzella, A. and Pizzuti, L. and Plana, C. and Polenta, G. and Pollack, J. E. and Poncet, M. and Pöntinen, M. and Pool, P. and Popa, L. A. and Popa, V. and Popp, J. and Porciani, C. and Porth, L. and Potter, D. and Poulain, M. and Pourtsidou, A. and Pozzetti, L. and Prandoni, I. and Pratt, G. W. and Prezelus, S. and Prieto, E. and Pugno, A. and Quai, S. and Quilley, L. and Racca, G. D. and Raccanelli, A. and Rácz, G. and Radinović, S. and Radovich, M. and Ragagnin, A. and Ragnit, U. and Raison, F. and {Ramos-Chernenko}, N. and Ranc, C. and Raylet, N. and Rebolo, R. and Refregier, A. and Reimberg, P. and Reiprich, T. H. and Renk, F. and Renzi, A. and Retre, J. and Revaz, Y. and Reylé, C. and Reynolds, L. and Rhodes, J. and Ricci, F. and Ricci, M. and Riccio, G. and Ricken, S. O. and Rissanen, S. and Risso, I. and Rix, H.-W. and Robin, A. C. and {Rocca-Volmerange}, B. and Rocci, P.-F. and Rodenhuis, M. and Rodighiero, G. and Monroy, M. Rodriguez and Rollins, R. P. and Romanello, M. and Roman, J. and Romelli, E. and {Romero-Gomez}, M. and Roncarelli, M. and Rosati, P. and Rosset, C. and Rossetti, E. and Roster, W. and Rottgering, H. J. A. and {Rozas-Fernández}, A. and Ruane, K. and {Rubino-Martin}, J. A. and Rudolph, A. and Ruppin, F. and Rusholme, B. and Sacquegna, S. and {Sáez-Casares}, I. and Saga, S. and Saglia, R. and Sahlén, M. and Saifollahi, T. and Sakr, Z. and Salvalaggio, J. and Salvaterra, R. and Salvati, L. and Salvato, M. and Salvignol, J.-C. and Sánchez, A. G. and Sanchez, E. and Sanders, D. B. and Sapone, D. and Saponara, M. and Sarpa, E. and Sarron, F. and Sartori, S. and Sassolas, B. and Sauniere, L. and Sauvage, M. and Sawicki, M. and Scaramella, R. and Scarlata, C. and Scharré, L. and Schaye, J. and Schewtschenko, J. A. and Schindler, J.-T. and Schinnerer, E. and Schirmer, M. and Schmidt, F. and Schmidt, F. and Schmidt, M. and Schneider, A. and Schneider, M. and Schneider, P. and Schöneberg, N. and Schrabback, T. and Schultheis, M. and Schulz, S. and Schwartz, J. and Sciotti, D. and Scodeggio, M. and Scognamiglio, D. and Scott, D. and Scottez, V. and Secroun, A. and Sefusatti, E. and Seidel, G. and Seiffert, M. and Sellentin, E. and Selwood, M. and Semboloni, E. and Sereno, M. and Serjeant, S. and Serrano, S. and Shankar, F. and Sharples, R. M. and Short, A. and Shulevski, A. and Shuntov, M. and Sias, M. and Sikkema, G. and Silvestri, A. and Simon, P. and Sirignano, C. and Sirri, G. and Skottfelt, J. and Slezak, E. and Sluse, D. and Smith, G. P. and Smith, L. C. and Smith, R. E. and Smit, S. J. A. and Soldano, F. and Solheim, B. G. B. and Sorce, J. G. and Sorrenti, F. and Soubrie, E. and Spinoglio, L. and Mancini, A. Spurio and Stadel, J. and Stagnaro, L. and Stanco, L. and Stanford, S. A. and Starck, J.-L. and Stassi, P. and Steinwagner, J. and Stern, D. and Stone, C. and Strada, P. and Strafella, F. and Stramaccioni, D. and Surace, C. and Sureau, F. and Suyu, S. H. and Swindells, I. and Szafraniec, M. and Szapudi, I. and Taamoli, S. and Talia, M. and {Tallada-Crespí}, P. and Tanidis, K. and Tao, C. and Tarrío, P. and Tavagnacco, D. and Taylor, A. N. and Taylor, J. E. and Taylor, P. L. and Teixeira, E. M. and Tenti, M. and Idiago, P. Teodoro and Teplitz, H. I. and Tereno, I. and Tessore, N. and Testa, V. and Testera, G. and Tewes, M. and Teyssier, R. and Theret, N. and Thizy, C. and Thomas, P. D. and Toba, Y. and Toft, S. and {Toledo-Moreo}, R. and Tolstoy, E. and Tommasi, E. and Torbaniuk, O. and Torradeflot, F. and Tortora, C. and Tosi, S. and Tosti, S. and Trifoglio, M. and Troja, A. and Trombetti, T. and Tronconi, A. and Tsedrik, M. and Tsyganov, A. and Tucci, M. and Tutusaus, I. and Uhlemann, C. and Ulivi, L. and Urbano, M. and Vacher, L. and Vaillon, L. and Valdes, I. and Valentijn, E. A. and Valenziano, L. and Valieri, C. and Valiviita, J. and den Broeck, M. Van and Vassallo, T. and Vavrek, R. and Venemans, B. and Venhola, A. and Ventura, S. and Kleijn, G. Verdoes and Vergani, D. and Verma, A. and Vernizzi, F. and Veropalumbo, A. and Verza, G. and Vescovi, C. and Vibert, D. and Viel, M. and Vielzeuf, P. and Viglione, C. and Viitanen, A. and {Villaescusa-Navarro}, F. and Vinciguerra, S. and Visticot, F. and Voggel, K. and {von Wietersheim-Kramsta}, M. and Vriend, W. J. and Wachter, S. and Walmsley, M. and Walth, G. and Walton, D. M. and Walton, N. A. and Wander, M. and Wang, L. and Wang, Y. and Weaver, J. R. and Weller, J. and Whalen, D. J. and Wiesmann, M. and Wilde, J. and Williams, O. R. and Winther, H.-A. and Wittje, A. and Wong, J. H. W. and Wright, A. H. and Yankelevich, V. and Yeung, H. W. and Youles, S. and Yung, L. Y. A. and Zacchei, A. and Zalesky, L. and Zamorani, G. and Vitorelli, A. Zamorano and Marc, M. Zanoni and Zennaro, M. and Zerbi, F. M. and Zinchenko, I. A. and Zoubian, J. and Zucca, E. and Zumalacarregui, M.},
  year = {2024},
  month = may,
  number = {arXiv:2405.13491},
  eprint = {2405.13491},
  primaryclass = {astro-ph},
  publisher = {arXiv},
  doi = {10.48550/arXiv.2405.13491},
  archiveprefix = {arXiv}
}

@misc{breivik2022,
  title = {From {{Data}} to {{Software}} to {{Science}} with the {{Rubin Observatory LSST}}},
  author = {Breivik, Katelyn and Connolly, Andrew J. and Ford, K. E. Saavik and Jurić, Mario and Mandelbaum, Rachel and Miller, Adam A. and Norman, Dara and Olsen, Knut and O'Mullane, William and {Price-Whelan}, Adrian and Sacco, Timothy and Sokoloski, J. L. and Villar, Ashley and Acquaviva, Viviana and Ahumada, Tomas and AlSayyad, Yusra and Alves, Catarina S. and Andreoni, Igor and Anguita, Timo and Best, Henry J. and Bianco, Federica B. and Bonito, Rosaria and Bradshaw, Andrew and Burke, Colin J. and {Rodrigues de Campos}, Andresa and Cantiello, Matteo and Caplar, Neven and Chandler, Colin Orion and Chan, James and {Nicolaci da Costa}, Luiz and Danieli, Shany and Davenport, James R. A. and Fabbian, Giulio and Fagin, Joshua and Gagliano, Alexander and Gall, Christa and Garavito Camargo, Nicolás and Gawiser, Eric and Gezari, Suvi and Gomboc, Andreja and {Gonzalez-Morales}, Alma X. and Graham, Matthew J. and Gschwend, Julia and Guy, Leanne P. and Holman, Matthew J. and Hsieh, Henry H. and Hundertmark, Markus and Ilić, Dragana and Ishida, Emille E. O. and Jurkić, Tomislav and Kannawadi, Arun and Kosakowski, Alekzander and Kovačević, Andjelka B. and Kubica, Jeremy and Lanusse, François and Lazar, Ilin and Levine, W. Garrett and Li, Xiaolong and Lu, Jing and Luna, Gerardo Juan Manuel and Mahabal, Ashish A. and Malz, Alex I. and Mao, Yao-Yuan and Medan, Ilija and Moeyens, Joachim and Nikolić, Mladen and Nikutta, Robert and O'Dowd, Matt and Olsen, Charlotte and Pearson, Sarah and Villicana Pedraza, Ilhuiyolitzin and Popinchalk, Mark and Popović, Luka C. and Pritchard, Tyler A. and Quint, Bruno C. and Radović, Viktor and Ragosta, Fabio and Riccio, Gabriele and Riley, Alexander H. and Rożek, Agata and {Sánchez-Sáez}, Paula and Sarro, Luis M. and Saunders, Clare and Savić, Đorđe V. and Schmidt, Samuel and Scott, Adam and Shirley, Raphael and Smotherman, Hayden R. and Stetzler, Steven and {Storey-Fisher}, Kate and Street, Rachel A. and Trilling, David E. and Tsapras, Yiannis and Ustamujic, Sabina and {van Velzen}, Sjoert and {Vázquez-Mata}, José Antonio and Venuti, Laura and Wyatt, Samuel and Yu, Weixiang and Zabludoff, Ann},
  year = {2022},
  month = aug,
  journal = {arXiv e-prints}
}

@inproceedings{racca2016,
  title = {The {{Euclid}} Mission Design},
  author = {Racca, Giuseppe D. and Laureijs, Rene and Stagnaro, Luca and Salvignol, Jean Christophe and Alvarez, Jose Lorenzo and Criado, Gonzalo Saavedra and Venancio, Luis Gaspar and Short, Alex and Strada, Paolo and Boenke, Tobias and Colombo, Cyril and Calvi, Adriano and Maiorano, Elena and Piersanti, Osvaldo and Prezelus, Sylvain and Rosato, Pierluigi and Pinel, Jacques and Rozemeijer, Hans and Lesna, Valentina and Musi, Paolo and Sias, Marco and Anselmi, Alberto and Cazaubiel, Vincent and Vaillon, Ludovic and Mellier, Yannick and Amiaux, Jerome and Berthe, Michel and Sauvage, Marc and Azzollini, Ruyman and Cropper, Mark and Pottinger, Sabrina and Jahnke, Knud and Ealet, Anne and Maciaszek, Thierry and Pasian, Fabio and Zacchei, Andrea and Scaramella, Roberto and Hoar, John and Kohley, Ralf and Vavrek, Roland and Rudolph, Andreas and Schmidt, Micha},
  year = {2016},
  month = jul,
  eprint = {1610.05508},
  primaryclass = {astro-ph},
  pages = {99040O},
  doi = {10.1117/12.2230762},
  archiveprefix = {arXiv}
}

@article{mclure2012,
  title = {The Sizes, Masses and Specific Star-Formation Rates of Massive Galaxies at 1.3},
  shorttitle = {The Sizes, Masses and Specific Star-Formation Rates of Massive Galaxies at 1.3},
  author = {McLure, R. J. and Pearce, H. J. and Dunlop, J. S. and Cirasuolo, M. and {Curtis-Lake}, E. and Bruce, V. A. and Caputi, K. and Almaini, O. and Bonfield, D. G. and Bradshaw, E. J. and Buitrago, F. and Chuter, R. and Foucaud, S. and Hartley, W. G. and Jarvis, M. J.},
  year = {2012},
  month = may,
  doi = {10.1093/mnras/sts092},
  language = {en}
}

@misc{jones2022,
  title = {Photometric {{Redshifts}} for {{Cosmology}}: {{Improving Accuracy}} and {{Uncertainty Estimates Using Bayesian Neural Networks}}},
  shorttitle = {Photometric {{Redshifts}} for {{Cosmology}}},
  author = {Jones, Evan and Do, Tuan and Boscoe, Bernie and Wan, Yujie and Nguyen, Zooey and Singal, Jack},
  year = {2022},
  month = feb,
  number = {arXiv:2202.07121},
  eprint = {2202.07121},
  primaryclass = {astro-ph},
  publisher = {arXiv},
  doi = {10.48550/arXiv.2202.07121},
  archiveprefix = {arXiv}
}

@article{garilli2014,
  title = {The {{VIMOS Public Extragalactic Survey}} ({{VIPERS}}): {{First Data Release}} of 57 204 Spectroscopic Measurements},
  shorttitle = {The {{VIMOS Public Extragalactic Survey}} ({{VIPERS}})},
  author = {Garilli, B. and Guzzo, L. and Scodeggio, M. and Bolzonella, M. and Abbas, U. and Adami, C. and Arnouts, S. and Bel, J. and Bottini, D. and Branchini, E. and Cappi, A. and Coupon, J. and Cucciati, O. and Davidzon, I. and De Lucia, G. and {de la Torre}, S. and Franzetti, P. and Fritz, A. and Fumana, M. and Granett, B. R. and Ilbert, O. and Iovino, A. and Krywult, J. and Le Brun, V. and Le Fèvre, O. and Maccagni, D. and Małek, K. and Marulli, F. and McCracken, H. J. and Paioro, L. and Polletta, M. and Pollo, A. and Schlagenhaufer, H. and Tasca, L. A. M. and Tojeiro, R. and Vergani, D. and Zamorani, G. and Zanichelli, A. and Burden, A. and Di Porto, C. and Marchetti, A. and Marinoni, C. and Mellier, Y. and Moscardini, L. and Nichol, R. C. and Peacock, J. A. and Percival, W. J. and Phleps, S. and Wolk, M.},
  year = {2014},
  month = feb,
  journal = {Astronomy \& Astrophysics},
  volume = {562},
  pages = {A23},
  issn = {0004-6361, 1432-0746},
  doi = {10.1051/0004-6361/201322790},
  language = {en}
}

@article{coil2011,
  title = {The {{PRIsm MUlti-Object Survey}} ({{PRIMUS}}) {{I}}: {{Survey Overview}} and {{Characteristics}}},
  shorttitle = {The {{PRIsm MUlti-Object Survey}} ({{PRIMUS}}) {{I}}},
  author = {Coil, Alison L. and Blanton, Michael R. and Burles, Scott M. and Cool, Richard J. and Eisenstein, Daniel J. and Moustakas, John and Wong, Kenneth C. and Zhu, Guangtun and Aird, James and Bernstein, Rebecca A. and Bolton, Adam S. and Hogg, David W.},
  year = {2011},
  month = nov,
  journal = {The Astrophysical Journal},
  volume = {741},
  number = {1},
  eprint = {1011.4307},
  primaryclass = {astro-ph},
  pages = {8},
  issn = {0004-637X, 1538-4357},
  doi = {10.1088/0004-637X/741/1/8},
  archiveprefix = {arXiv}
}

@article{newman2013,
  title = {The {{DEEP2 Galaxy Redshift Survey}}: {{Design}}, {{Observations}}, {{Data Reduction}}, and {{Redshifts}}},
  shorttitle = {The {{DEEP2 Galaxy Redshift Survey}}},
  author = {Newman, Jeffrey A. and Cooper, Michael C. and Davis, Marc and Faber, S. M. and Coil, Alison L. and Guhathakurta, Puragra and Koo, David C. and Phillips, Andrew C. and Conroy, Charlie and Dutton, Aaron A. and Finkbeiner, Douglas P. and Gerke, Brian F. and Rosario, David J. and Weiner, Benjamin J. and Willmer, Christopher N. A. and Yan, Renbin and Harker, Justin J. and Kassin, Susan A. and Konidaris, Nicholas P. and Lai, Kamson and Madgwick, Darren S. and Noeske, Kai G. and Wirth, Gregory D. and Connolly, Andrew J. and Kaiser, Nick and Kirby, Evan N. and Lemaux, Brian C. and Lin, Lihwai and Lotz, Jennifer M. and Luppino, Gerard A. and Marinoni, Christian and Matthews, Daniel J. and Metevier, Anne and Schiavon, Ricardo P.},
  year = {2013},
  month = aug,
  journal = {The Astrophysical Journal Supplement Series},
  volume = {208},
  number = {1},
  eprint = {1203.3192},
  primaryclass = {astro-ph},
  pages = {5},
  issn = {0067-0049, 1538-4365},
  doi = {10.1088/0067-0049/208/1/5},
  archiveprefix = {arXiv}
}

@article{cool2013,
  title = {The {{PRIsm MUlti-object Survey}} ({{PRIMUS}}). {{II}}. {{Data Reduction}} and {{Redshift Fitting}}},
  author = {Cool, Richard J. and Moustakas, John and Blanton, Michael R. and Burles, Scott M. and Coil, Alison L. and Eisenstein, Daniel J. and Wong, Kenneth C. and Zhu, Guangtun and Aird, James and Bernstein, Rebecca A. and Bolton, Adam S. and Hogg, David W. and Mendez, Alexander J.},
  year = {2013},
  month = apr,
  journal = {The Astrophysical Journal},
  volume = {767},
  number = {2},
  eprint = {1303.2672},
  primaryclass = {astro-ph},
  pages = {118},
  issn = {0004-637X, 1538-4357},
  doi = {10.1088/0004-637X/767/2/118},
  archiveprefix = {arXiv}
}


\appendix

\section{Data Creation}
\label{section:dataset}
\subsection{COSMOS2020 Photometric Redshifts}
\label{section:cosmos2020}
In this work, we make use of photometric redshifts from the latest release of the Cosmic Evolution Survey (COSMOS; \cite{scoville2007}) catalog (COSMOS2020; \cite{weaver2022}) consisting of over 1 million sources. The COSMOS field covers about 1.7 deg$^{2}$ of the sky and has been observed across the electomagnetic spectrum by the Galaxy Evolution Explorer (GALEX; \cite{zamojski2007}) in the far-UV to near-UV, the Canada-France-Hawaii Telescope Large Area U-band Deep Survey (CLAUDS; \cite{sawicki2019}), the Hyper Suprime-Cam Subaru Strategic Program (HSC-SSP; \cite{aihara2019}) and Suprime-Cam (SC) data (\cite{taniguchi2007, taniguchi2015}) on the Subaru telescope in the optical, the UltraVISTA survey (UVISTA; \cite{mccracken2012, milvang-jensen2013}) in the near-infrared, and the Cosmic Dawn Survey \cite{sanders2007, moneti2022} using Spitzer in the mid-infrared. The surveys cover the area in the X-ray, optical, and infrared, enhancing our studies in galaxy evolution and nature of dark matter. 

The COSMOS2020 catalog is composed of two separate catalogs labeled \texttt{CLASSIC} and \texttt{THE FARMER}. \texttt{CLASSIC} uses aperture photometry methods while \texttt{THE FARMER} uses profile fitting methods (The Tractor;\cite{lang2016}) for the photometry measurements of extended sources. 
In addition, the catalogs provide photometric redshifts from two independent template fitting codes \texttt{LePhare} \cite{arnouts1999} and \texttt{EAZY} \cite{brammer2008} along with additional outputs. 

In this work, we make use of sources that are common to both catalogs. The \texttt{CLASSIC} catalog contains $1, 720, 700$ sources with photometric redshifts computed using up to 35-bands while \texttt{THE FARMER} consists of $964, 506$ sources with photometric redshifts computed using up to 30-bands. The discrepancy arises from the variable Point Spread Function (PSF) of the Suprime-Cam medium bands, which \texttt{THE FARMER} could not overcome. After cross-matching using internal IDs, there is $923, 403$ common sources among the two catalogs. 

Our decisions in Section \ref{section:transferz} depend on the reliability of the photometric redshifts. Among the different combinations of photometry (\texttt{CLASSIC}/\texttt{THE FARMER}) and photometric redshift codes (\texttt{LePhare}/\texttt{EAZY}), the numbers of bands fit by the photo-z codes is not consistent. \texttt{CLASSIC}/\texttt{LePhare} uses up to 35 bands. \texttt{CLASSIC}/\texttt{EAZY} uses up to 30 bands, excluding GALEX data and Suprime-Cam broad bands due to their limited depth and PSF issues (Suprime-Cam and UltraVISTA narrow bands are also excluded, though no explicit reason is provided). \texttt{THE FARMER}/\texttt{CLASSIC} uses up to 30 bands, excluding the Suprime-Cam broad bands. \texttt{THE FARMER}/\texttt{EAZY} uses up to 27 bands, excluding the Suprime-Cam broad bands, GALEX FUV/NUV bands, and all narrow bands. While photometry cannot reach the precision of spectroscopy, more photometric bands reduces degeneracies in redshift measurements. Since \texttt{CLASSIC}/\texttt{LePhare}'s photometric redshifts are estimated using the largest number of bands, this set is chosen as the labels used in our neural network training involving TransferZ. Specifically, we use \texttt{lp\_zPDF} for a galaxy corresponding to the median of the photometric redshift likelihood distribution.

\subsection{HSC Photometry}

For our analysis we compile five-band (\textit{grizy}) photometry to approximate data produced by large scale surveys in comparable depth \cite{ivezic2008, breivik2022, euclidcollaboration2024, racca2016}. We use data from the HSC Subaru Strategic Program's (HSC-SSP) second data release (HSC PDR2; \cite{aihara2019}) in the wide field that reaches similar depths as LSST but over a smaller area coverage. The HSC wide-field camera is mounted on the Subaru Telescope with a FOV of 1.8 deg$^2$. The survey has observed over 900 deg$^2$ of the sky across five optical filters (\textit{grizy}) with a median seeing in the i-band of 0.58" and a median 5$\sigma$ depth of 26.2. The HSC-SSP survey does not have a bluer band unlike LSST which is expected to have six optical filters \textit{ugrizy}.    

Our analysis draws from a query of over 3 million sources around the COSMOS field from the HSC-PDR2 Wide catalogs. Our photometry data is queried from HSC-SSP's data release site. No initial constraints are placed on this query of our data. In addition, we query for a sample of galaxies with spectroscopic redshifts for validation. In Section \ref{section:transferz}, we use these spectroscopic redshifts to guide the quality of our cuts.

\subsection{TransferZ}
\label{section:transferz}
\begin{figure}[htb]
    \centering
    \includegraphics[width=5.5 in]{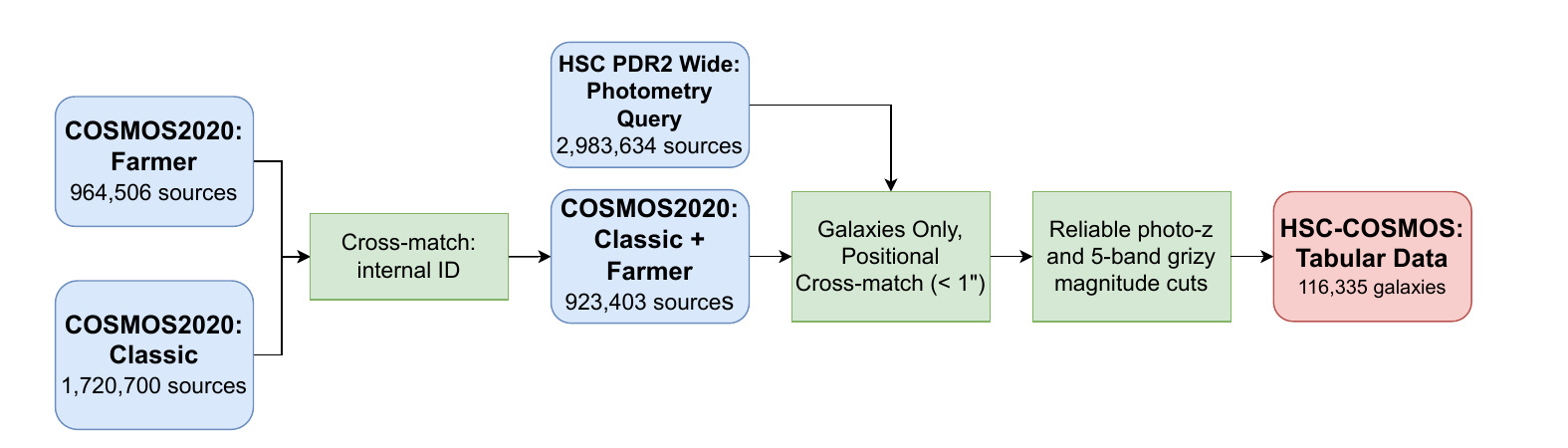}
    \caption{Flow chart showing the steps used in creating the TransferZ dataset. Green rectangles represent processes and blue rectangles represent inputs. The red rectangle is the dataset released with this paper.}
    \label{fig:transferz_flow}
\end{figure}

We created the TransferZ dataset, a dataset consisting of galaxies with \textit{grizy} photometry and reliable photometric redshifts following the workflow shown in Figure \ref{fig:transferz_flow}. The creation of TransferZ is split into two main steps: combining galaxy HSC data with COSMOS2020 data and performing quality cuts to ensure reliable features and labels. In this work, HSC \textit{grizy} photometry serve as our features and \texttt{LePhare} photometric redshift \texttt{lp\_zPDF} from the \texttt{CLASSIC} catalog of COSMOS2020 serve as our labels. See Section 
\ref{section:cosmos2020} for more details about these criteria. 

\subsubsection{Cross-match Datasets and Galaxy Filter}
The combination of HSC and COSMOS2020 data involved two steps. (1) A positional cross-match of HSC PDR2 sources with the COSMOS2020 catalog subset from Section \ref{section:cosmos2020} within 1". (2) We filter for galaxies using \texttt{LePhare} classification method \texttt{lp\_type = 0} when available, otherwise defaulting to HSC source classification \texttt{i\_extendedness\_value = 1}. We are aware that \cite{weaver2022} advises against using \texttt{LePhare} classifcation; however, we find a strong agreement between HSC and COSMOS2020 classification methods, with 88\% accuracy when we consider COSMOS2020 classification as the ground truth. The step reduces our dataset to 670,053 galaxies. 

\subsubsection{Quality Cuts}
The quality cuts are similar to \cite{singal2022} where the cuts are split in two categories: COSMOS2020 photometry cuts and direct cuts to the photometric redshift quality. We note that the analysis in their work makes use of COSMOS2015 \cite{laigle2016}, a previous iteration of the COSMOS catalog which consists of aperture based photometry and photometric redshifts computed by \texttt{LePhare} photo-z code. 

First, we implement cuts on COSMOS2020 photometry quality used in their redshift estimation. We require quality measurements in the bands used for their redshift estimation, as poor quality or missing photometry degrades estimated redshifts. We limit the requirement of photometry between 0 and 50 mag for 23 bands from HSC (\textit{grizy}), Suprime-Cam (medium band filters), UVISTA (\textit{YJHKs}), and \textit{Spitzer}/IRAC (channel 1 and channel 2). These filters are consistent across estimated photometric redshifts from the four configurations (aperture based and profile-fitting based photometry processed by \texttt{LePhare} and \texttt{EAZY}). The u and u$^*$ band filters are ignored as galaxies at redshift $z \approx 3.3$ and beyond would not be detected in these bands due to absorption by neutral hydrogen (this absorption creates what is known as the Lyman break at a rest-frame wavelength of 912 \AA).

Second, we apply cuts to ensure photometric redshift quality and agreement between all redshift configurations in COSMOS2020. We first exclude galaxies lacking redshift measurement across all configuration and require robust $\chi^2$ fits in both photo-z codes. Following \cite{singal2022}, we applied a threshold of $\chi^2 < 1$ for fits from \texttt{CLASSIC}/\texttt{LePhare}. For the other configuration, we define the threshold to remove the same fraction of galaxies as the \texttt{CLASSIC}/\texttt{LePhare} threshold. Additionally, for \texttt{LePhare} redshift estimates, we require close agreement ($|$\texttt{lp\_zPDF} - \texttt{lp\_zMinChi2}$| < 0.1$) between the redshift at the peak of the likelihood distribution and the redshift that minimizes $\chi^2$.  We also require \texttt{CLASSIC}/\texttt{LePhare} photometric redshifts to agree with the other configurations, as \texttt{CLASSIC}/\texttt{LePhare} photometric redshifts serve as the ground truth labels in our training. We restrict our sample to $z<4$. 

Finally, we implement cuts on our training features which we use the HSC-PDR2 \textit{grizy} photometry \texttt{cModel} magnitudes. We require these values to be between 0 and 50 mag. After all cuts, TransferZ contains 116,335 galaxies.

\end{document}